# Harnessing constrained resources in service industry via video analytics

Chun-Hung Cheng, Iyiola E. Olatunji *
Department of Systems Engineering and Engineering Management
The Chinese University of Hong Kong
Shatin, Hong Kong
(chcheng@se.cuhk.edu.hk, iyiola@link.cuhk.edu.hk)
* Corresponding Author

*Abstract*— Service industries contribute significantly to many developed and developing - economies. As their business activities expand rapidly, many service companies struggle to maintain customer's satisfaction due to sluggish service response caused by resource shortages. Anticipating resource shortages and proffering solutions before they happen is an effective way of reducing the adverse effect on operations. However, this proactive approach is very expensive in terms of capacity and labor costs. Many companies fall into productivity conundrum as they fail to find sufficient strong arguments to justify the cost of a new technology yet cannot afford not to invest in new technologies to match up with competitors. The question is whether there is an innovative solution to maximally utilize available resources and drastically reduce the effect that the shortages of resources may cause yet achieving high level of service quality at a low cost.

This work demonstrates with a practical analysis of a trolley tracking system we designed and deployed at Hong Kong International Airport (HKIA) on how video analytics helps achieve management's goal of satisfying customer's needs via real-time detection and prevention of problems they may encounter during the service consumption process using existing video technology rather than adopting new technologies. This paper presents the integration of commercial video surveillance system with deep learning algorithms for video analytics. We show that our system can provide accurate decision when faced with total or partial occlusion with high accuracy and it significantly improves daily operation. It is envisioned that this work will heighten the appreciation of integrative technologies for resource management within the service industries and as a measure for real-time customer assistance.

## I. INTRODUCTION

In Service industries, their operations are not just a tool for revenue generation but an integral part of organization's success with responsiveness as an important factor for service receiver's satisfaction [1]. Responsiveness refers to how fast a customer can get a required service or the provider's response time to receiver's request while maintaining high level of service quality. Service quality is the comparison between actual and expected service provided within a reasonable tolerance margin. In this work, organizations, service providers or providers are used interchangeably which refers to the providers or sources of a service. Also customers, service receivers or receivers refer to the consumers of the service.

Service can be anything from taking orders, providing support, responding to customer's request (customer service) to provisioning of resources needed for customer's satisfaction. Resource refers to material or tool that has the prospect of creating value when used and operated upon [2]. Therefore, resource may range from value, money, capacity, assets, brand, experiences, staff, materials, equipment or their combinations.

Every organization is constrained in resources. However, under-harnessing the available resources and infrastructure often result in poor service operations. The service receiver's level of participation in different services highly influences the type of service provided [3]. When the level of participation is high, service receiver co-creates the service i.e. availability of receiver highly influences the service provided while when the level of participation is low, services are provided irrespective of whether there is an immediate receiver to consume the service. An example of high level of service receiver's participation is in personal training service offered in most fitness centers and gyms. Since most people have different objectives in their fitness goal, a tailored made fitness plan is created for each individual. In this type of service, the personal trainer liaises with the individual and they both workout a fitness plan that help the individual achieve the desired fitness goal. On the other hand, trolley availability in an airport operation is an example of service provided with low customer participation. In such a case, trolleys must always be available at strategic locations regardless of whether an immediate traveler (service receiver) is available to use it or not due to the difference in people inflow caused by varying flight schedule.

The effectiveness of staffing decisions which usually account for a significant portion of organization's operational cost is measured by the ratio of staffing levels to revenues [4]. However, such measurement does not explicitly consider non-staff contribution in the generation of revenues for situations in which service receiver's participation is low. This motivates the development of an efficient and fast method to measure customer assistance in real time through the use of analytic result derived from video data.

Capacity management is a complex and difficult task encountered by service organizations with little information to assist them. Considering the nature of services being perishable and accounting for an increasing percentage of the gross output in most countries [5], an appropriate proportion of demand to supply has been peddled as one of the major problems faced by managements in service industries. The primary capacity resource in service supply chain (SSC) is skilled labor which incurs a lot of cost [6]. A puzzling question that follows is "how can we reduce labor cost while keeping an adequate level of service quality?". Therefore, organization's capacity has to be managed optimally for maximum utilization and effective productivity. Despite its importance, we observe that there has been lack of attention devoted to using existing technologies to achieve maximum capacity utilization and resource management.

Facility‐based service organizations and field‐based service organizations are the two major categories of service organizations [7]. In facility-based service organizations, customers are saddled with the responsibility of accessing the service facility while in field-based service organizations, the organization or service provider is obliged to provide services to service receiver at receiver's site. In our research, we only consider facility-based service organizations as capacity management is a major problem in these organizations.

According to [8], innovation is one of two basic functions of an organization to gain and retain competitiveness in the current business environment. Innovative behavior has a positive effect on organization's growth. Therefore, the significance of innovation and creativity for business competitiveness and need for change to cater for shifting needs of an organization cannot be overemphasized in the current knowledge-based economy. With the service industry attracting attention in world economy [9], innovative techniques to foster development of new services is required.

However, many organizations fall into productivity conundrum in which they are in pursuit of new technologies to sustain their competitiveness yet fail to find sufficient justification for the new technologies [10]. This motivates our work on using existing technologies as a solution to the productivity conundrum for resource management. The value of an innovation is defined by how firms can better serve instead of the organization's output [2], [11] which is achieved by integrating existing and sometimes new resources in innovative ways.

Therefore, the aim of this work is to integrate existing technologies, more specifically video streams from CCTV with deep learning models to effectively manage capacity resource demand and supply rather than adopting new technologies which are costlier and may require long adoption time. We also explore how video analytic as a form of innovation can drive change in the service industry

Knowledge is a central resource in service industries and its economic utilization is the measure of organization growth. Data like services are intangible and generating knowledge from data requires a framework or predefined structure for efficient distribution since the generation or acquisition of data may come from different sources [12]. Analyzing video data can be used to effectively utilize available resources and has the capability of avoiding poor service planning through the use of powerful analytics tools to transform visual contents into knowledge. It offers valuable way of expressing observed scenes in complex environment where variability, uncertainty, and inaccuracy must be taken into account [13].

Conventionally, the fundamental practice of a provider is to produce, while that of receiver is to consume [14], [15]. Analyzing visual data for resource management makes the production process timely and swift thereby creating more opportunities to enact different practices created by the service. Agnihothri et al. [7] showed that effective strategic choices are made by managers guided by an integrated framework compared to managers not guided by frameworks.

The application of video technology into service industries has the potential of radically transforming existing service processes and driving new ones. However, the impact of video technology in service industry has been insufficiently examined as most studies focus on self-service automation [16], business‐to‐business relationships [17] and product sales [18]. This paper aims to bridge the gap through exploration of how video technology is changing service processes with practical applications.

The objective of this paper is to examine the changing nature of resource management with an emphasis on how video analytics can be used for resource management by turning videos into tracking data and actionable intelligence for easy decision making. The contributions of our work are in three folds:

1. To drive innovative change in the service industry through the use of existing technology at a low cost using video analytics (solution to productivity conundrum).

2. We present with a practical trolley management system we designed and implemented at Hong Kong international Airport (HKIA), the use of integrative technologies for capacity and resource management in service industry.

3. We analyze the framework of the system for resource management and as a measure for real time customer assistance.

The rest of this paper is organized as follows: First, we review the current methods of resource management using a pilot project of different resource tracking techniques we conducted at HKIA and how it affects service provisioning. We discuss advantages of video analytics in service industry and provide extensive review on methods of resource monitoring from service industry point of view and resource management point of view. In section 3, we describe how video analytics can be integrated with a practical deployment of trolley tracking system in a typical airport operation. Discussion on the method and the advantages of using videos over existing resource management technology are analyzed in section 4. Section 5 concludes the paper.

## II. LITERATURE REVIEW

Over the years, new opportunities for service innovation have been birth especially those driven digitally [19] which has dramatically transformed many organizations. Traditional approaches to managing an enterprise are becoming obsolete due to the e-business revolution thereby prompting organizations to review their methods in order to stay competitive and manage daily task effectively and seamlessly [20]. Systems regarded as enterprise resource planning (ERP) systems have been successful in the re-engineering of these traditional management approaches. ERP involves a suite of software for the collection, storage, management and interpretation of business data. However, input from video streams for resource management may not be easily captured by this system due to the complexity of video data.

Similarly, powerful knowledge management systems have been adopted by organizations due to the recent advances and maturity of technologies in order to maintain or create sustainable competitive advantage [21]. Organization's objectives are usually hardwired into knowledge management systems such as performance improvement, innovations and competitive advantage with the intention of creating, sharing and acting upon the knowledge for effective decision making and actualization of organization goals [22]. This ensures that organizations have up-to-date knowledge on products, and services. This increases efficiency, saves time and prevents duplication of work due to availability of knowledge but is relatively expensive.

The goal of service offerings is to strengthen credibility of the organization and to improve customer's satisfaction [23] which can be made possible via informational resources such as those derived from video analytics. Video technology allows for high flexibility of customization as it is suitable for expressing scenes in complex environment where variability, uncertainty, and inaccuracy is constant e.g. pooled service. Pooled service is a type of service interdependence that allows the allocation of majority of the work to back-office operation which are entirely separated from front–office influence and independent of the customers [24]. Remote monitoring of CCTV camera is an example of this type of service.

Traditional conceptualization of resource management from high-touch, low-tech to low-touch, high-tech paradigm [25] is theatrically changing through the introduction of technology where the availability of resource influences the quality of service provided. Consider a service of trolley provisioning for passengers in an airport operation depicted by two scenarios. Scenario 1: a customer service representative is deployed strategically to communicate with floor staff on deployment of trolleys (resource) to certain areas of the airport based on passenger's inflow. Scenario 2: a video camera monitors the availability of the trolley (resource) and alerts floor staff accordingly. In the first scenario, replenishment of resources may not be effective i.e. service responsiveness will be quite slow and costly due to labor cost. On the other hand, the second approach uses existing infrastructure such as the CCTV to provide additional information and efficient resource management strategy. These two scenarios illustrate the traditional and technological driven approach of fulfilling the same service respectively.

In addition to traditional service industries such as hotel, airline and banks, product-based firms (non-service industries) are gradually offering services as a product which has proven to be more profitable and serve as key strategy for competitive marketplace constrained by availability of resources [25]. The role of video analytics in service industries provides substantial benefits to both service providers and service receivers. Service providers can monitor their resources easily via comprehensive data mining of video data represented quantitatively while service receivers benefit via responsiveness to their need.

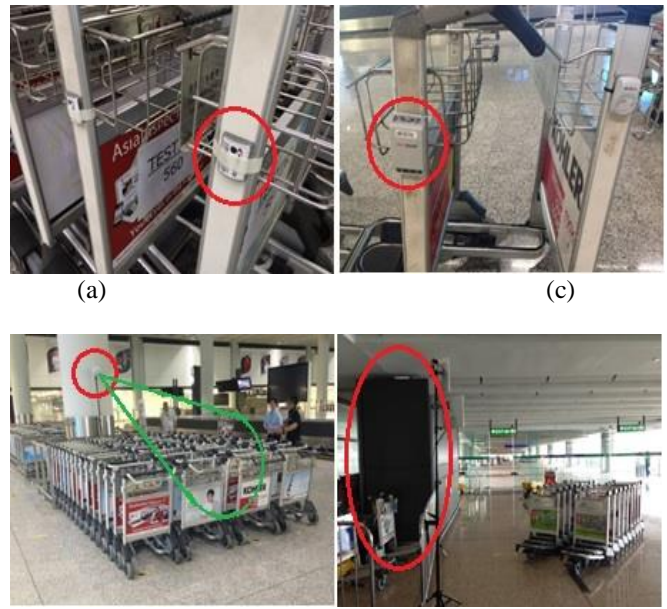

Figure 1. Pilot study of different trolley tracking solutions conducted at Hong Kong International Airport (HKIA). (a) – (b): RFID solution. The RFID tag is attached to each trolley and an RFID reader is mounted to receive signals of each trolley. (c) – (d): WiFi Solution. WiFi signal tag is attached to each trolley and a reader to receive WiFi Signal. Red circles are the transmitter and receivers for each solution respectively

The human-resources trap described by [26] as organizations excessively focusing on delivery of service by human and interpersonal contact while neglecting other non-personal ways of interacting with customers is one motivation for our work. Real time adaptability to customer's need is advantageous for service providers and considering the limitations of humans, using videos for resource management for effective provisioning of service plays an important role. This type of innovation provides flexibility, accuracy and speed and often enhances the value of customer's experience. Transitioning of visual data to quantitative data can provide knowledgeable insights and key information which can be effectively used by service providers to enhance decision making.

Timely production and consumption of service where and when needed has proven to be important factors for satisfaction, thus automation is indispensable. Due to the importance of service automation in service provisioning [24], this work provides an enlightenment into how video technology enables new service processes.

The increasing complexity of algorithms and growth of camera installation have made video surveillance intelligent spanning over several fields from networks and sensors to algorithms and software development [27]. However, the software components still need to be customized for the desired need. Several research has been conducted for software modules such as motion detection, action recognition, object recognition [28]-[31] but practical integration with existing systems is still a grey area. Video analytics refers to software tools that provides quantitative representation of visual data from observed scene composed of a suite of pattern recognition and computer vision modules for knowledge extraction from detected visual data [27].

Localization technologies such as barcode, radio frequency identification (RFID) and Bluetooth technologies are widely used in the service industry for resource management. Barcode technology have been popularly applied to retail stores for tracking of products. Furthermore, a DNA barcoding enterprise for taxonomy has been proposed as a service industry in which barcode technology is used in identification and classification of genetically modified organism (GMO) and contaminants to non-taxonomist [32]. RFID provides an excellent tracking for resources [33], however they are quite expensive both at the initial procurement as well as the maintenance. For example, periodic check of the RFID tag must be performed on the resources in order to provide accurate resource monitoring which incurs additional operational cost. Bluetooth technology offers robust low power consumption and low complexity solution [34]. Bluetooth has been used in hospitals for real time collection of personal health data and synchronized with hospital information system for bio-sensing data generation [35] but requires the receiver to be actively involved in the service generation process which does not fit into our low service receiver participation application. For example, the service receiver must constantly wear a device to for the service throughout service process.

As shown in figure 1, pilot study of using RFID and WiFi based solutions was conducted at HKIA. However, both approaches have physical limitation of tag placement on trolleys, battery cost and tag replacement when they are faulty. Signal receiver (reader) are physically constrained by space (WiFi solution) or lower accuracy due to trolley detection in semicircle for RFID solution (green circle).

A similar paper [34] to our work that used video methodologies in service industry. However, they used motion segmentation in multimedia surveillance to improve RFID localization rather than in resource management which makes our work one of the first studies in the application of video methodologies for resource management. In [34], combination of RFID and motion segmentation was used to determine the physical location of a moving object in indoor environment. Determining the physical location of moving objects is important in surveillance system where precise location cannot be easily derived from GPS data. The integration of RFID technology with image processing techniques including object recognition, tracking and localization of moving object, and block matching algorithm were used for analyzing the image for motion segmentation and object tracking.

Other applications of video-based methodologies in service systems are in capturing real-time perceptions of customer processes [36] and navigating multi-modal public transport systems [37]. Video‑based methodology was used in the collection of real-time user perception data for determining important factors for measurement and analysis of service processes but they both focus on customer's experience. In [38], indoor video tracking technique was used in the collection of shopping behavior dataset. Each shopper wore a portable Bluetooth-sized camera. Their shopping behavior from the product category consideration to point of purchase was captured coupled with entrance and exit survey of their shopping plan. These were used for the generation of shopping behavior dataset. Knowledge generation or identification, transfer, storage and application are the major challenges of knowledge management in knowledge-intensive services [39].

Since it will be challenging and costly for humans to monitor huge volumes of generated video data, processing video streams on the go is required for current surveillance systems. Object detection and tracking are the fundamentals of video processing techniques [40]. For example, to recognize an activity from a video stream, it depends on detecting an object and tracking the object throughout the lifespan of the video stream.

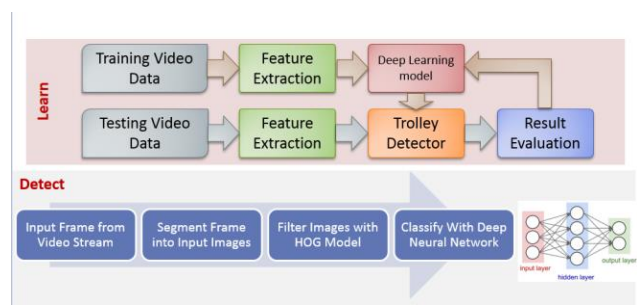

Figure 2. Simplified learning and detection phase of the trolley tracking system

Efficiency is about providing services that maximally utilize infrastructure, reliable, interpretable, understandable and are readily available to all parties involved in the service production and consumption. Using video analytics for resource management maximizes efficiency and it allows several questions to be easily answered such as capacity utility maximization, current available resources, people inflow and interaction with the resource.

## III. INTEGRATIVE TECHNOLOGY FOR RESOURCE MANAGEMENT

According to [41], the resource configuration that makes services unique and of significant contribution to the service success is the methods and tools. Therefore, careful configuration of the techniques and tools is required. In this section, we examined the method used in the development of our trolley tracking system and how they are integrated into the existing system of HKIA.

### A. Methodology and Design

Due to occlusion, clustered background and viewpoint variation, object recognition using video surveillance is a challenging task in real world environment. Machine learning algorithms, more specifically deep neural networks models have shown powerful capabilities in analyzing complex scenes from video data. Deep learning models refers to a class of machine learning models that can learn features from raw data where multiple layers learn multiple level of abstraction.

A three-layer hierarchical method [28] was used to solve the problem of multiple target tracking in non-overlapping cameras using constrained dominant sets. The first two layers were used for in-camera tracking while the third layer was used for tracking across camera. Previous approach on video prediction involves reconstruction of future frames from the internal state of the model. However in [29], object and background appearance are not stored but rather merged with motion predicted by the model which allows prediction of future video sequences for multiple steps. Our tracking model framework is inspired by the framework of [28], [29] to provide video frame prediction as well as object detection for better decision making.

Our trolley tracking model uses Histogram of oriented gradient (HOG) as feature extractor and fed into a deep neural network for trolley identification. The region of interest is dynamically mapped using background subtraction and texture analysis. The convolution layer of our network combines feature map with several convolution kernels and produces output feature map equivalent to the number of input feature map as represented in equation 1.

$$y = \max\left(0, \sum_{k} f_k * h_k + b\right) \quad (eqn\ 1.)$$

$$y_{(i,j)} = \max_{0 \leq m,n \leq s} \{x_{(i \cdot s + m, j \cdot s + n)}\} \quad (eqn\ 2.)$$

$$y = \max(0, x) \quad (eqn\ 3.)$$

$$A = \frac{TTP}{AGT} \quad (eqn\ 4.)$$

$$F = \frac{TFP}{TTP + TFP} \quad (eqn\ 5.)$$

*where f is the input feature map of the kth frame, h is the convolutional kernel, ∗ is the convolution operation and b is the bias term. s is the pooling size, x represents the convolution output, TTP is the total number of true positives, AGT is the total number of actual ground truth and TFP is the total number of false positives.*

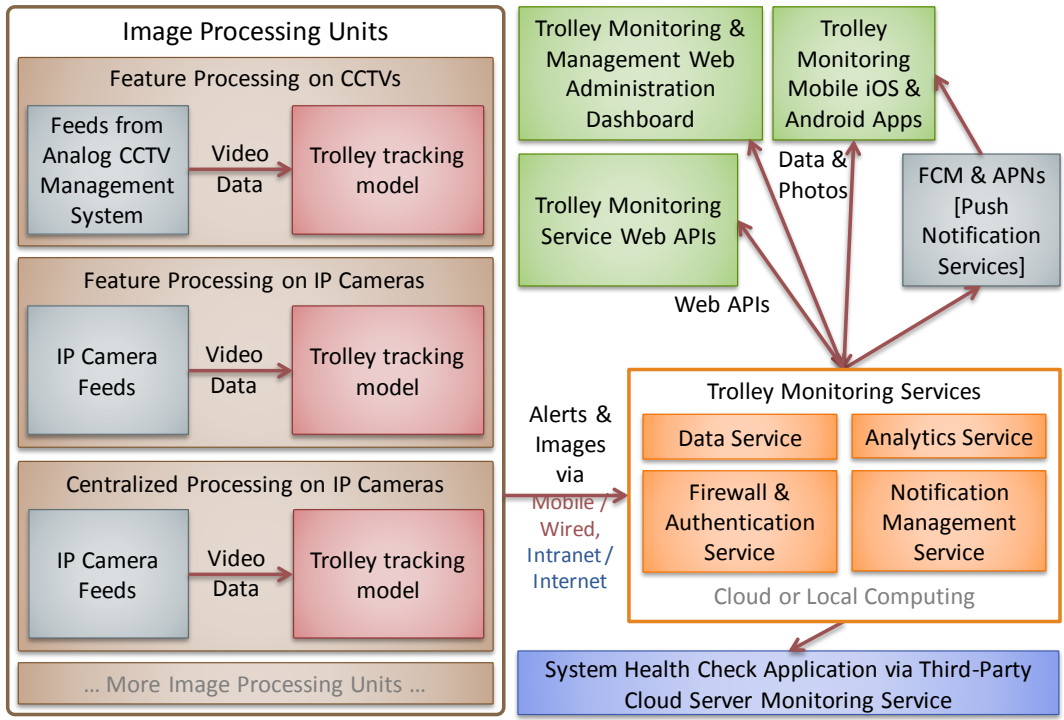

Figure 3. Trolley tracking system framework

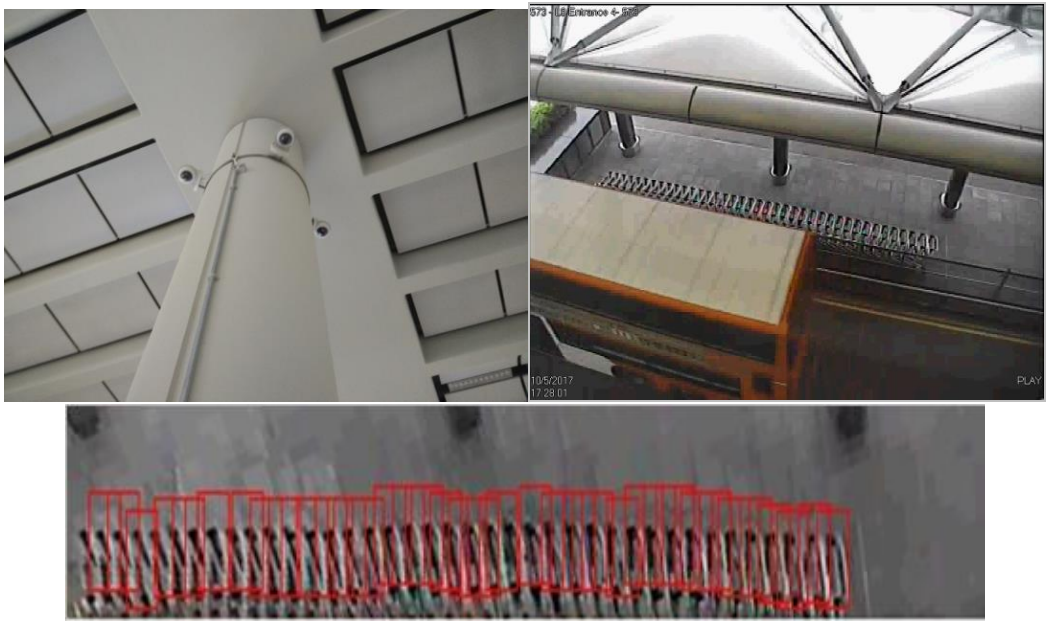

Figure 4. Real time video stream processing from CCTV camera installation with a vehicle obstructing camera view. The picture with red bounding

Both the convolutional kernel and bias term are continuously learned in the training process. We used the rectified linear unit (ReLu) (equation 3) for non-linear transformation mapping of the input to output for each convolution. The max-pooling layer (equation 2) is used intermittently after each convolution to select feature subset. The performance of our system is evaluated using accuracy and false alarm rate as shown in equation 4 and 5 respectively. Intricate details of the system architecture such as number of convolutional layers, hyper parameters settings etc. are not discussed in this paper due to the recent commercialization of our system.

### B. Trolley Tracking System

Baggage trolley as we know it is an important resource in an airport operation. For example; according to the air traffic statistics of Hong Kong International Airport (HKIA) [42], over 60 million travelers were served by the HKIA in 2017 with around 13,000 baggage trolleys distributed throughout the airport. Coping with this large demand due to enormous passenger flow and limited resources requires a well-defined architecture to effectively manage this resource (baggage trolley) while providing high level of trolley availability to travelers.

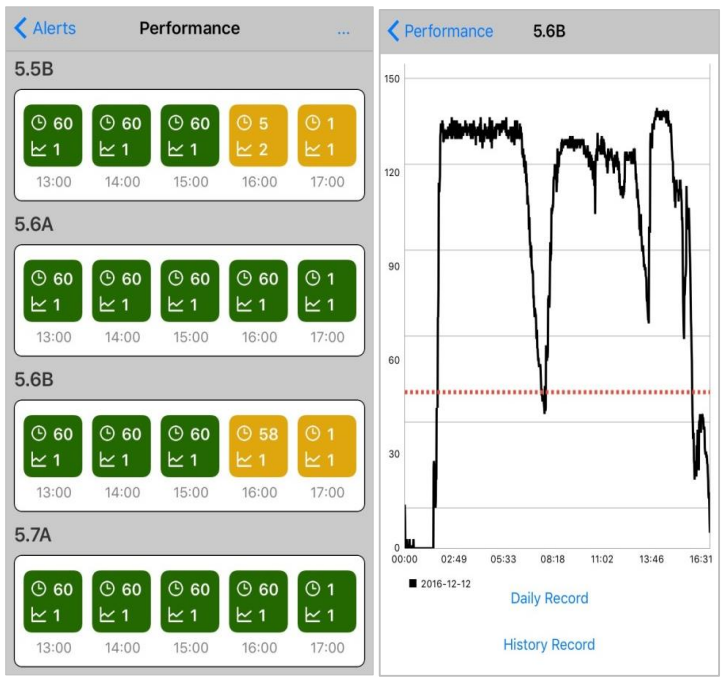

Fig 5a. Trolley tracking availability display and statistics on mobile App. Green, yellow and red color coding signifies good, warning and critical respectively

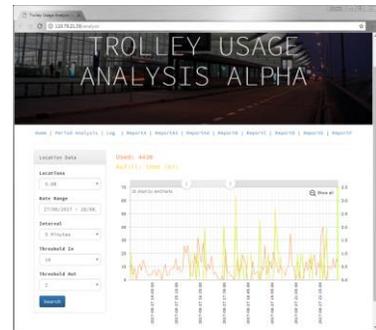

Fig 5b. Web API integration into HKIA existing management system

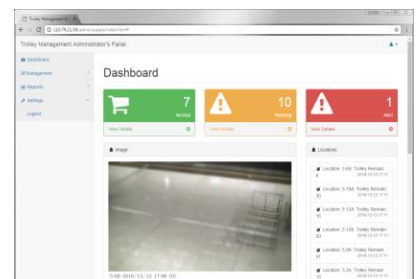

Fig 5c. Analytics and reporting

As shown in Figure 3, the trolley tracking system consists of two (2) modules: Image processing module and Trolley monitoring module. The image processing unit takes raw CCTV camera video which serves as input for the trolley tracking model. The trolley tracking model consists of the identification submodule and counting submodule. The identification submodule distinguishes other objects in the video stream from the intended trolley object while blurring out other objects such as humans for privacy and security purposes. The input into the counting module is the correctly identified trolleys with blur on other objects of each video frame.

The counting submodule keeps track of the available numbers of trolley. Figure 4 shows a typical CCTV camera installation with real-time video stream into the trolley tracking model with partial occlusion. The identified trolleys are shown in red bounding box. When total occlusion occurs, the video stream is paused and restarted afterwards upon detection of trolleys.

service submodule (input from tracking model), analytics submodule (Data mining), firewall and authentication submodule and notification management submodule. Query endpoint of the monitoring module is provided via Web API, mobile App and Web App. The purpose of the Web API is to allow integration with existing system while the Web App and mobile App are full fledge implementation and display of analytic results. Figure 5b and 5c shows the Web API integration into HKIA existing system and Web app respectively. Display of analytic trend on mobile app is shown in Figure 5a. Green (good), yellow (warning), red (critical) are color codes used to depicts the status of available trolleys. When there is decrease in the available trolleys or at certain threshold, a notification is sent for replenishment of trolley in that particular location with information containing other location with less trolley usage. Our trolley tracking system has been deployed in the baggage reclaim hall, departure hall and taxi drop station at Hong Kong International Airport (HKIA).

IV. MANAGERIAL IMPLICATION AND SYSTEM EVALUATION

The significance of our trolley tracking system for effective capacity management, daily business operations and to drive customer's satisfaction is invaluable. Our trolley tracking system has made significant improvement to the

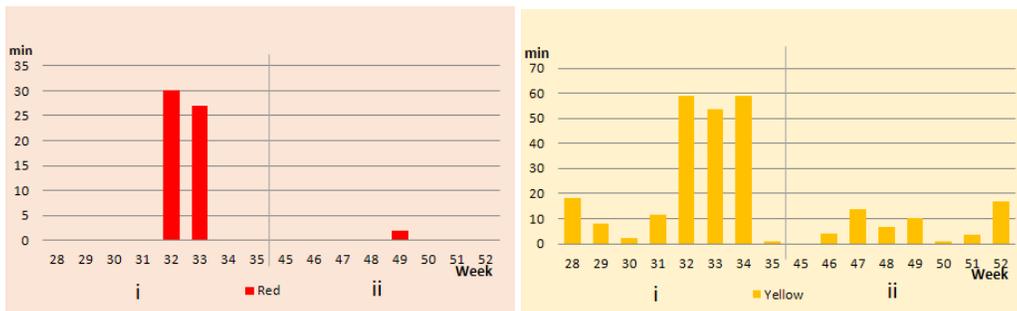

Figure 6a. Weekly cumulative time of staying in critical (red) and warning (yellow) alert status within initial week of deployment (i) and latter weeks of deployment (ii)

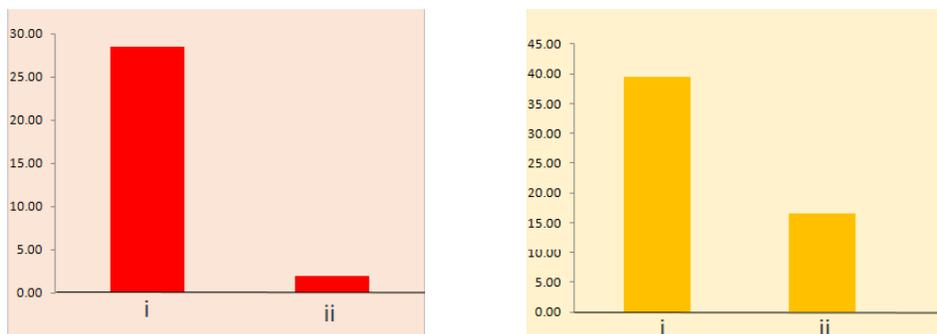

Figure 6b: Average trolley replenishment response time for critical (red) and warning (yellow) alert status within initial week of deployment (i) and latter weeks of deployment (ii)

The trolley monitoring module converts the processed video data into quantitative data so as to provide knowledgeable insights via analytics. It consists of data availability and management of trolleys deployed in over 18 stations in the baggage reclaim hall. Trolley replenishment time for trolley stations at critical alert status have significantly reduced from almost 30 minutes to 2 minutes (~

93% reduction) as shown in Figure 6b. Likewise, in figure 6a, the weekly cumulative time of a trolley station staying at critical level was also reduced by over 10 times the magnitude at the latter weeks of deployment (~ 97% reduction). Our system has an overall accuracy of 91.85% as shown in figure 8. Figure 8 ascertains the robustness of our system in detecting trolleys even with partial occlusion. Pointless to say the response time before the deployment of our trolley tracking system was extremely high. The reason for the drastic reduction in replenishment time is due to the learning curve of using a new system and as such is independent of our system performance. Trolley contractor's manpower of patrolling has significantly reduced since the need for going to each trolley station to manually monitor usage has been eliminated via our mobile analytic platform and thus assist contractor for better trolley management performance.

Contract management between trolley contractor and the airport authority can be enhanced by our analytic system by monitoring usage of the trolleys as well as understanding the dynamics of fast moving trolley stations as shown in figure 7. Besides labor cost and resource usage understanding, no maintenance cost is required compared to RFID and WiFi resource management solution. Chain of reaction caused by incessant gathering at one or multiple stations due to resource shortage can be predicted and eradicated. Also, collection of current status at different locations can be easily streamlined thereby enabling an integrated monitoring of transit center.

V. CONCLUSION

This paper presents with a practical stellar example of the use of video analytics in service industry for resource management using existing video infrastructure. This leads to low cost resource tracking solution. Since real-time replenishment of resources is desired by both service provider and service consumer, constrained resources can be maximally utilized via our proposed framework. In this paper, we showed the integration of video streams from CCTV with deep learning as a solution to productivity conundrum faced by most service industries. We also discussed managerial implication based on the system deployment in Hong Kong International Airport.

Experimental result shows that with partial or total occlusion, our system can still provide accurate decision making via analytics of the video stream without raising false alarms.

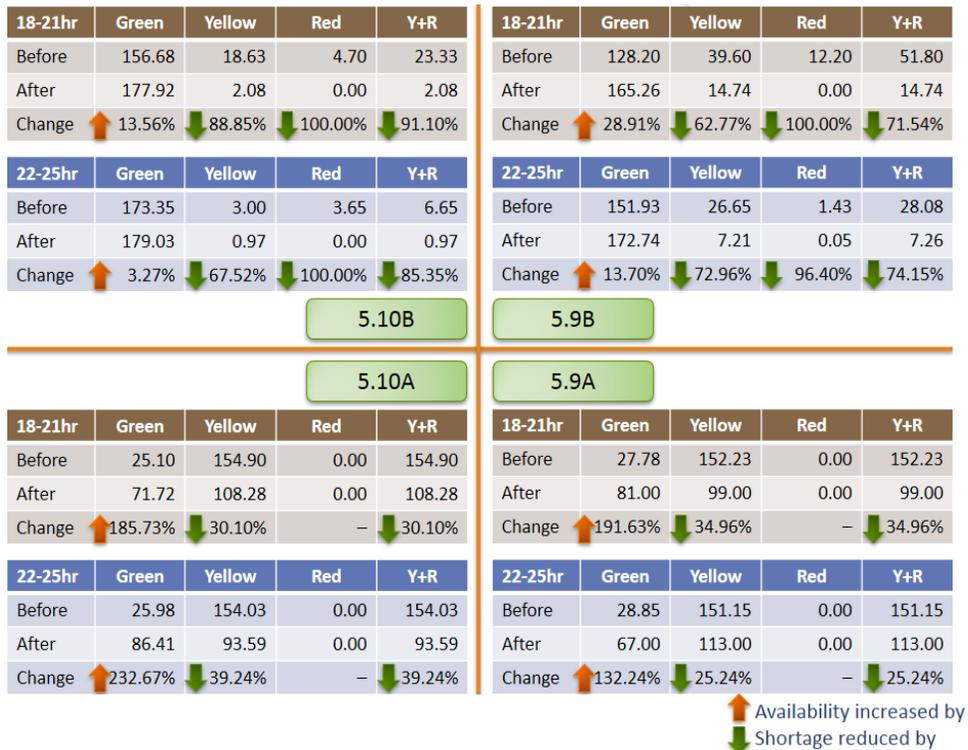

Fig 7. Comparing performance differences in 4 different trolley stations

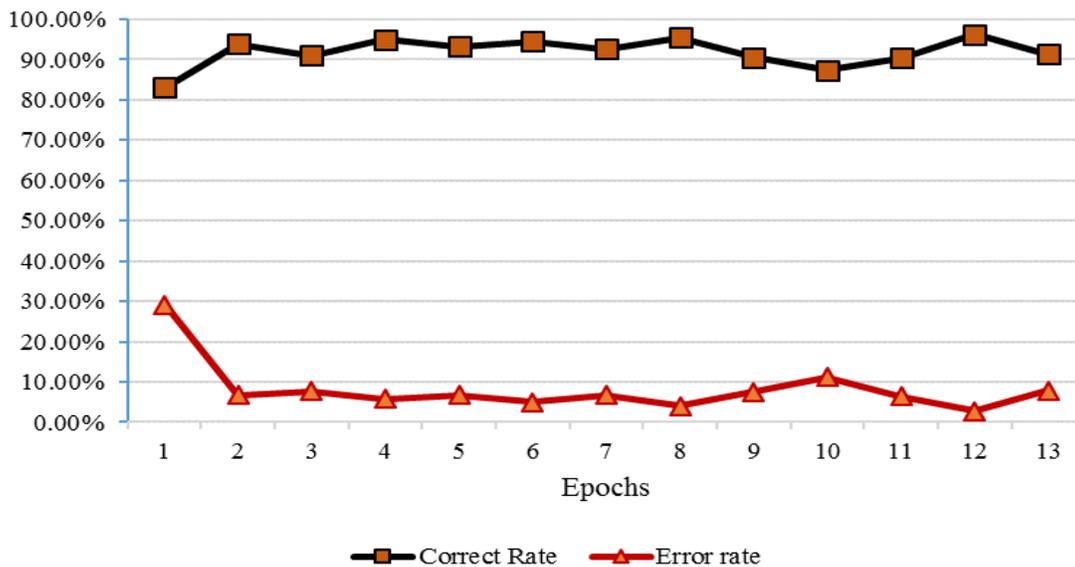

Fig 8. Counting submodule error rate vs accurate detection

ACKNOWLEDGMENT

This research has been partially supported by a GRF grant (Project No. 14615015) from Research Grants Council of Hong Kong SAR, China.